\documentclass{JINST}
\pdfoutput=1
\usepackage{graphicx}
\usepackage{url}
\usepackage{cite}
\usepackage{lineno}
\usepackage{subfigure}

\title{A Development Environment for Visual Physics Analysis}

\author{H.-P.~Bretz, M.~Brodski, M.~Erdmann, R.~Fischer, A.~Hinzmann, T.~Klimkovich, D.~Klingebiel, M.~Komm, J.~Lingemann, G.~M\"uller, T.~M\"unzer, M.~Rieger, J.~Steggemann, T.~Winchen\\
        RWTH Aachen University, Physikalisches Institut 3A, 52062 Aachen, Germany\\
        E-mail: \email{erdmann@physik.rwth-aachen.de}}

\abstract{
The Visual Physics Analysis (VISPA) project integrates different aspects of physics analyses into a graphical development environment.
It addresses the typical development cycle of (re-)designing, executing and verifying an analysis.
The project provides an extendable plug-in mechanism and includes plug-ins for designing the analysis flow, for running the analysis on batch systems, and for browsing the data content.
The corresponding plug-ins are based on an object-oriented toolkit for modular data analysis.
We introduce the main concepts of the project, describe the technical realization and demonstrate the functionality in example applications.}


\begin{document}


\section{Introduction}

Physics data analysis aims at extracting information from data taken by an experimental apparatus.
The analysis and understanding of the data is typically an interplay of designing (and implementing) algorithms and of interpretation of the results.
As the amount of data and the complexity of analyses tend to increase in modern experiments, the necessary amount of time for the technical integration of the individual analysis steps rises as well.
The intention of Visual Physics Analysis (VISPA) is to increase the available time for the interpretation of the results by reducing the time spent on the technical aspects of the data analysis.
The VISPA project provides a graphical development environment for physics analyses, including a selection of plug-ins specifically designed for the needs of data analyses in high energy physics (HEP) and astroparticle physics.
The target group ranges from undergraduate students to experienced scientists.
It has been continuously developed since 2006~\cite{vispa,vispaacat2011,vispaacat2010,vispaweb,vispachep2010,vispaepshep2009,vispaconcepts,vispanovel,pax06}.

In the field of high energy physics, a graphical analysis development environment constitutes a new approach.
Most high energy physics collaborations provide dedicated software frameworks for different tasks ranging from reconstruction to user analysis~\cite{cmsframework, athena, augerframework}.
While these frameworks attempt to provide solutions for multiple aspects of data processing, the VISPA environment is designed as a standalone application dedicated to high-level analysis with frequent development cycles using a reduced data format.
Visual support is provided for the development, execution and verification of analysis workflows.
Specific tasks like plotting, histogramming, and statistical analysis are not at the core of this development environment; they are, e.g.,\ accessible via the \verb+ROOT+ framework~\cite{root}.

We define a list of requirements for a development environment for physics analyses, which can be categorized into three groups.
The first group of requirements is due to the complexity of modern physics analyses.
A proper management of the analyses needs to be ensured by providing a structural basis, objects to represent an analysis, and
development tools to keep track of the analysis structure.
At the same time, the physicists should not be limited to predefined analysis techniques, conserving the freedom of scientists to create and develop new ideas.
Therefore, useful functionality should be provided by the development environment, without invoking restrictions on analysis techniques. 
Another important requirement is an acceptable amount of time for the iteration of the analysis cycle, which typically consists of an analysis design phase, program execution, data verification, and re-design of the analysis.

The second group of requirements is related to the wide range of programming skills of the analyzing physicists.
A fast start in physics analyses helps analyzers with little programming knowledge.
It is advantageous to learn through intuitive and self-explanatory design.
Appropriate guidance by the provided structures and interfaces as well as good transparency of the analysis
help to avoid errors from user code, which is important also for the experienced physicists.

The third group of requirements is driven by the fact that many physics analyses are carried out in teams.
The analysis development environment needs to support and facilitate this teamwork.
This implies that analyses need to be easily exchangeable between physicists and their individually preferred working environments and especially operating systems.
Visual representation of an analysis is desired to ease communication between analysts.
Code and algorithms should be reusable in other ongoing or future analyses.

This article continues with a description of the project structure and the design decisions to meet the requirements for a development environment for physics analyses.
The subsequent section gives an overview of the implementation of the VISPA graphical user interface.
Afterwards, the three main components needed to design physics analyses are explained, namely a collection of physics objects and algorithms,
a framework supporting modular physics analyses, and the graphical tools for physics analysis.
Finally, example applications and extensions are detailed, followed by the conclusions.

\section{Analysis Development Environment}

\begin{figure}[h!]
\begin{center}
\includegraphics[width=0.85\textwidth]{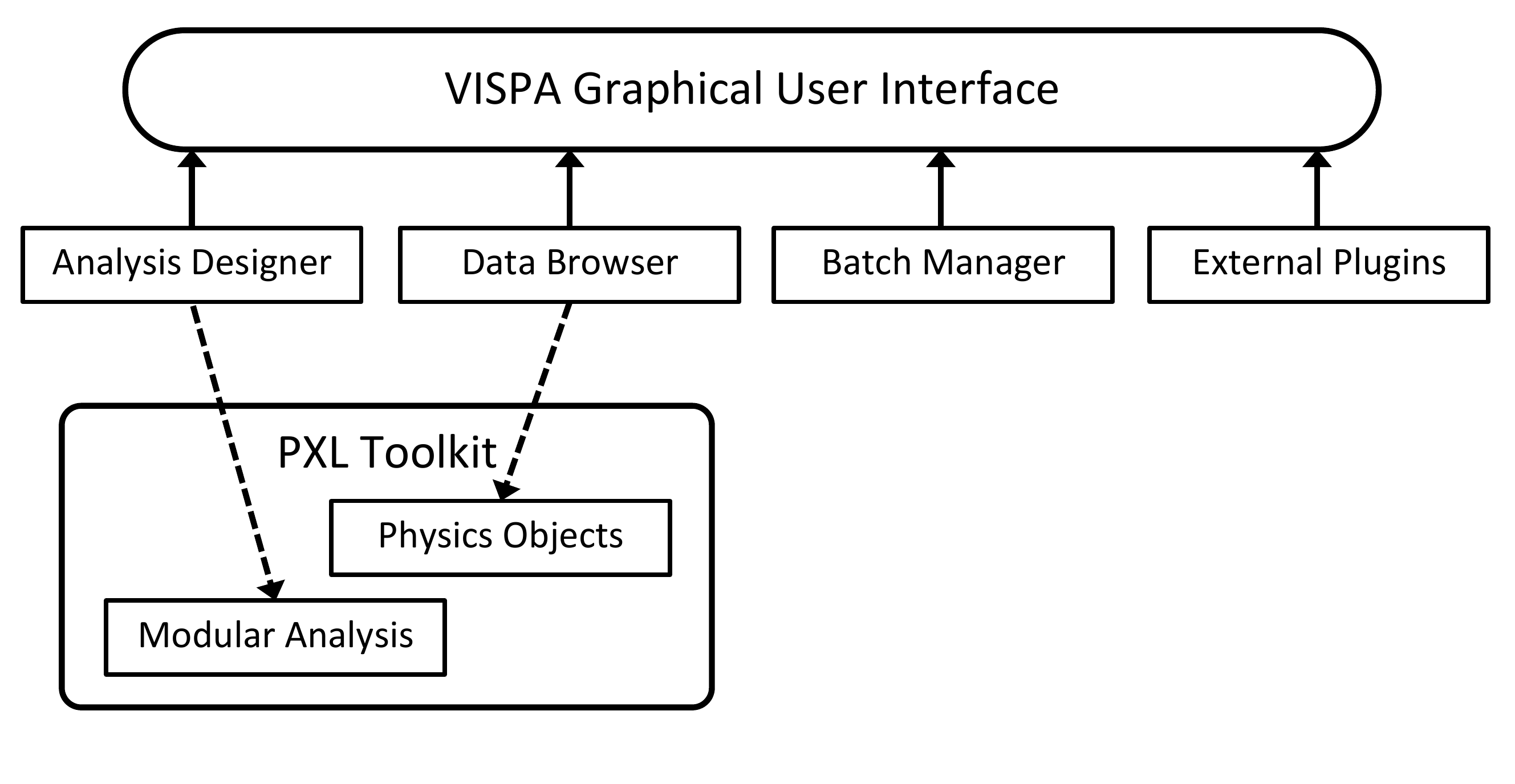}
\end{center}
\caption{Structure of the VISPA development environment.
The central entry point to physics analysis is the VISPA GUI.
Tools for physics analysis development are provided as plug-ins.
The analysis-specific plug-ins, namely the \emph{Analysis Designer} and the \emph{Data Browser}, are based on the functionalities of the PXL toolkit.
External plug-ins from collaborations and users can be plugged in the VISPA environment as well.}
\label{fig:vispastructure}
\end{figure}

The general structure of the VISPA development environment is depicted in Figure~\ref{fig:vispastructure}.
The central entry point to physics analysis is the VISPA graphical user interface (GUI).
Tools for physics analysis development are provided as plug-ins.
The three major plug-ins that allow for the complete handling of physics analyses are the \emph{Analysis Designer}, the \emph{Data Browser} and the \emph{Batch Manager}. With the \emph{Analysis Designer}, physics analyses can be designed and executed. The \emph{Data Browser} is a tool to browse the data input and output written in the file format of the Physics eXtension Library (PXL)~\cite{pxl}. Both the \emph{Analysis Designer} and the \emph{Data Browser} are based on PXL, which is a part of the VISPA project. PXL is a C++ toolkit that provides physics objects, data input and output, a module system, and other tools to program physics analyses. The \emph{Batch Manager} is able to configure batch jobs, in particular from analyses designed with the \emph{Analysis Designer}, and to send them to a selected batch system. Further tools, e.g.,\ browsers for experiment-specific workflows, can be added to the graphical development environment with the plug-in system.

To fulfill the requirements given in the introduction, several design choices build the basis for VISPA and are described in the following.
The first choice is bundling of the iterative analysis development process, consisting of analysis prototyping, execution and verification of the results, into a single development environment.
This integrated design gives rise to a good manageability of the full analysis cycle.

Second, instead of pre-defining analysis solutions by providing each single building block of an analysis, VISPA provides tools for the user to design an analysis structure.
The algorithms are input by the user or a collaboration.

Third, VISPA is designed to be extendable.
This is realized via the plug-in mechanism, allowing the inclusion of plug-ins that are based on the VISPA GUI.
By enabling experiment-specific extensions of the VISPA development environment, it is possible to handle the complete analysis workflow including steps involving software from the experiment in a single development environment.

Fourth, VISPA is designed to run on all major operating systems, i.e.\ Linux, Microsoft Windows and Mac OS X.
This feature is particularly important to enable the sharing of algorithms between users and the ability to transport entire analyses.
The VISPA GUI is therefore based on the platform-independent GUI framework \verb+PyQt4+~\cite{pyqt,qt}.

\section{Graphical User Interface}

\begin{figure}[h!]
\begin{center}
\includegraphics[width=0.53\textwidth]{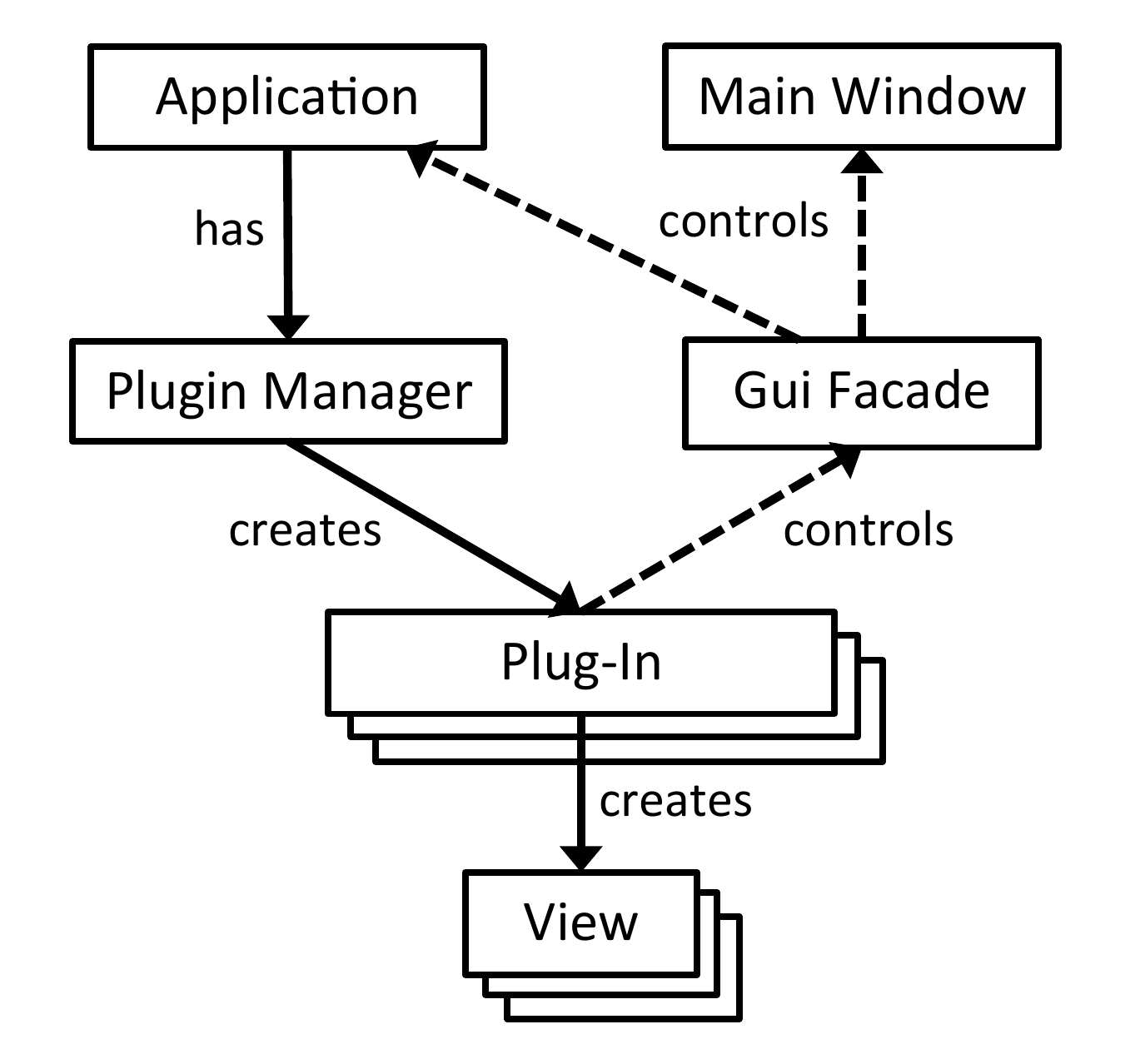}
\end{center}
\caption{Structure of the VISPA GUI implementation.
The full VISPA GUI class structure is documented in~\cite{vispaclasses}.}
\label{fig:vispagui}
\end{figure}

The core components of the VISPA GUI are the classes \emph{Application} and \emph{MainWindow} as depicted in Figure~\ref{fig:vispagui}.
They provide common functionality for opening and closing files, creating corresponding menu entries, and opening and closing tabs.
A single interface to this functionality for the plug-ins is provided by the class \emph{GuiFacade}.
On startup of the application, a \emph{PluginManager} searches for available plug-ins and creates the corresponding instances.
The plug-ins themselves can create new tabs via a singleton object of the class \emph{GuiFacade}, e.g.,\ on a file open request from the Application.

For the graphical display of objects within tabs, such as analysis modules or data objects, a mechanism for data-model driven \emph{Views} is introduced in VISPA.
By defining a common interface between data objects and views, a single view can be reused for different data objects.
In addition, the same data objects can be displayed in different views, e.g.,\ to represent different aspects of the object.
VISPA supplies a set of views covering a large variety of use cases in custom applications.
Experiment-specific views can also be implemented based on the existing views.

While the VISPA GUI is solely dependent on \verb+Python+ and \verb+PyQt+, custom views and plug-ins may dynamically use other external dependencies.
An example for a view using an external library is the \emph{RootCanvasView}, which is capable of displaying graphs produced with \verb+ROOT+, e.g.,\ the distribution of particle momenta in the angular plane.

For the development of new views, a variety of shared components is available.
These include useful graphics components, such as connectable boxes and lines, a common zooming mechanism for all components in VISPA and the functionality of each compound to be exported to various raster or vector graphic formats (e.g.\ \verb+PostScript+).

\begin{figure}[h!]
\begin{center}
\includegraphics[width=0.8\textwidth]{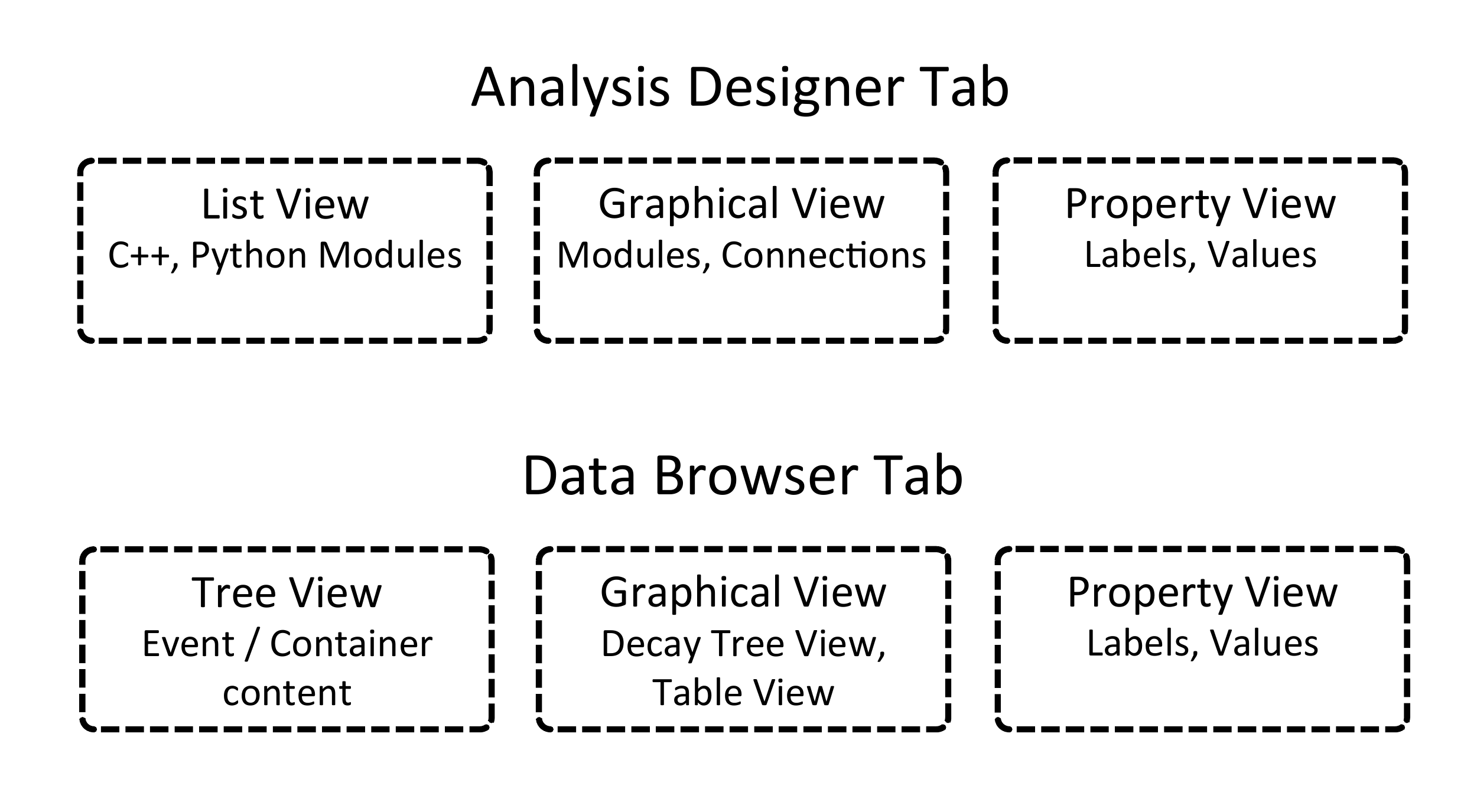}
\end{center}
\caption{Stylized layout of the \emph{Analysis Designer} and \emph{Data Browser} tabs.}
\label{fig:vispawindow}
\end{figure}

The VISPA GUI and its handling are designed with reoccurring patterns for good manageability and a fast understanding of the structure.
An example for a reoccurring pattern is the three-column representation of analyses in the \emph{Analysis Designer} and of data files in the \emph{Data Browser} as depicted in Figure~\ref{fig:vispawindow}.
In the \emph{Analysis Designer}, the left column presents available modules for analysis, whereas in the \emph{Data Browser}, the data objects in a file are displayed in a tree structure.
The center column holds a graphical representation of the analysis or the data, respectively.
Depending on the view, this could for instance be a Feynman-like representation.
The properties of an object selected in the center column are displayed in a \emph{Property View} in the right column.

\section{Physics Library}

In the VISPA environment, analyses are constructed based on classes representing physics objects and algorithms provided by the C++-based toolkit PXL, the successor of the PAX toolkit~\cite{pax06,pax,pax05,pax03, pax1}.
PXL offers a variety of classes for physics objects and algorithms as well as tools for code and analysis handling.
The individual components, described in the following, are designed to be used independently as tools for physics analysis or within the VISPA environment.

PXL contains classes to represent objects from high energy physics, from astroparticle physics and for general purposes.
An example from HEP is the class \emph{Particle} which contains a four-momentum and properties such as the particle charge.
For the field of astroparticle physics, e.g.,\ classes representing ultra-high energy cosmic rays (UHECR) are available together with common operations such as transformations between astrophysical coordinate systems.
General-purpose classes include matrix and vector representations.

\begin{figure}[h!]
\begin{center}
\includegraphics[width=0.65\textwidth]{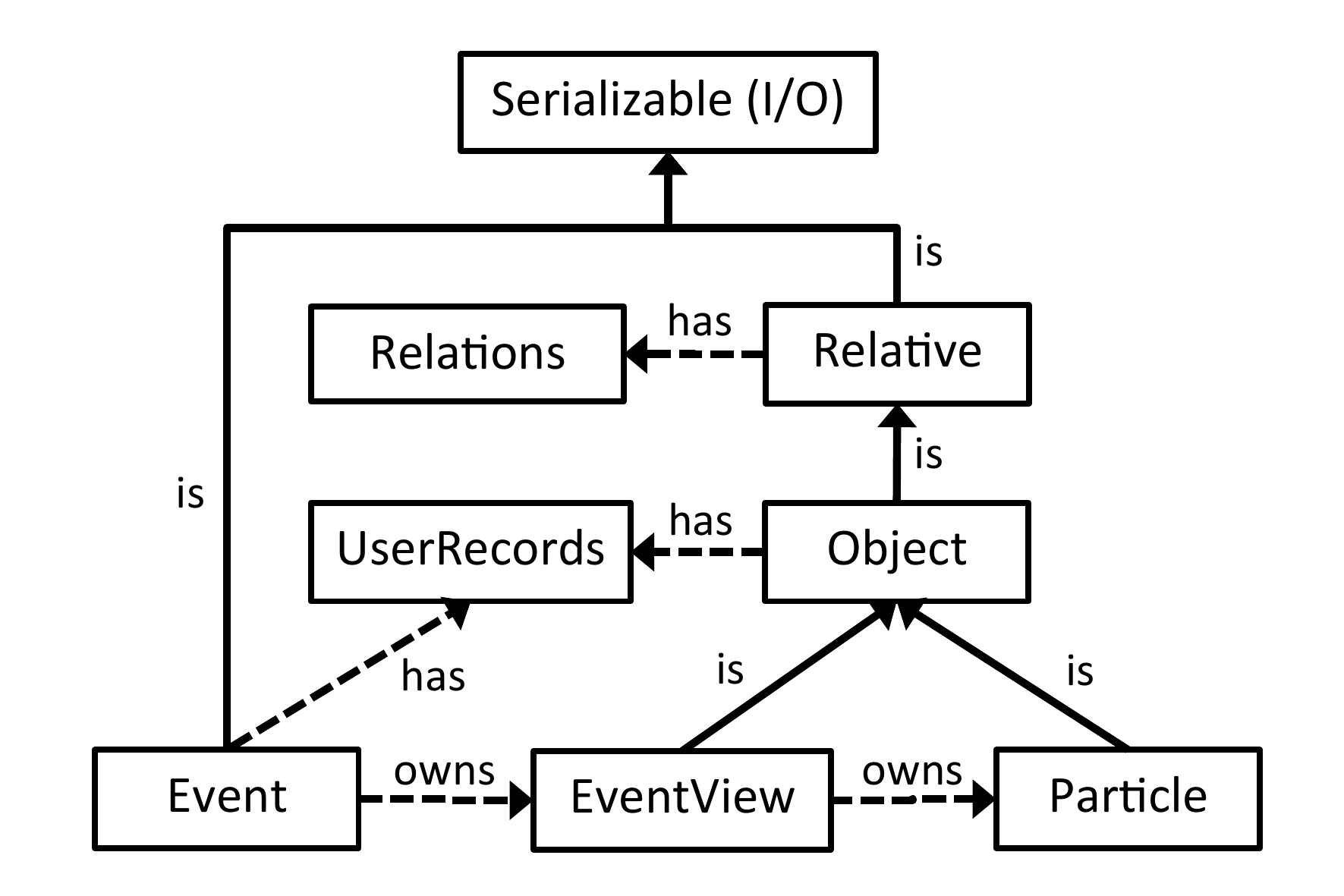}
\end{center}
\caption{Excerpt of the class structure of physics objects in PXL showing the structure of selected HEP objects.
The full PXL class structure is documented in~\cite{classref}.}
\label{fig:pxlobject}
\end{figure}

An excerpt of the class structure of selected HEP objects in PXL is shown in Figure~\ref{fig:pxlobject}.
For HEP, two base classes are provided from which the physics objects and user-defined objects can inherit.
The \emph{Object} provides user-specific data and
the \emph{Relative} can have relations with other objects, which are explained later in this section.
Similarly, base classes for astroparticle physics objects are provided.
All objects carry universally unique identifiers (UUIDs~\cite{uuid}) which are used to identify C++ classes (type ID) as well as each individual object (object ID).

PXL provides different types of containers to group various PXL objects.
The \emph{BasicContainer} is a collection of PXL objects that takes ownership and deletion responsibility for the inserted objects.
The \emph{Event} provides an extended container specifically designed for holding HEP objects.
An example are particles of a decay tree which have mother-daughter relations.
Deleting the \emph{Event} implies automated deletion of the particles contained in the event and their relations.
Furthermore, PXL provides containers which themselves can have relations with other objects.
A specific use case from HEP is an \emph{Event} holding a set of containers representing different views of the same physics event, so called \emph{EventViews}, e.g.,\ if there are ambiguities in the reconstruction leaving room for different interpretations.

A key feature in PXL to provide full flexibility in the implementation of analyses is the possibility to extend any object by user-specific data.
Each \emph{Object} has a \emph{UserRecords} instance which is a map of string-indexed entries of generic type.
Generic types are represented by a variant data type that covers basic C++ types, the STL string representation and any PXL object type.

Relations between physics objects can be realized in PXL through two different mechanisms depending on the use case.
Firstly, relations of objects within the same container (e.g.\ particles in an event) are realized in such a way that the
container takes care of the deletion of its relations between them to ensure full consistency at all times.
A \emph{Relative}, for example, has two \emph{Relations} objects to implement mother and daughter relations.
Secondly, so called soft relations are realized by a map of string-indexed relations to generic objects even outside a container through their unique object ID.

The PXL objects come with an input/output (I/O) scheme where each object is decomposed into serially written basic types.
All objects inheriting from the class \emph{Serializable} define how to (de-) serialize themselves, and can thus be used in the PXL I/O scheme.
The I/O is managed in a set of classes that provide the user interface for writing and reading files, gathering data chunks, managing compression, controlling the buffer storage and managing the basic type I/O.
PXL I/O specifically uses 32 or 64 bit variables to correctly support both systems.
All data is written to disc in little endian format, as this is most  commonly used.
On big endian systems all data is converted.
The PXL I/O aims for robustness and simplicity and is designed for simple splitting and merging of data at the file level.

To allow the use of C++ and Python code within the same analysis, the full C++ interface of all PXL classes
is made available in Python.
The Python interfaces are wrapped around the C++ classes automatically using \verb+Swig+~\cite{swig}.
Automatic conversion between C++ and Python types is also provided such that all PXL objects can be handled like native Python objects.
The Python interface of PXL allows full introspection of all properties and methods of any object and also supplies descriptions for all of these.
Full introspection of all objects in PXL enables the \emph{Data Browser} plug-in of VISPA to provide visual inspection of all properties of any PXL object.

For easier code and analysis handling, a set of convenience mechanisms is provided by PXL.
For example, a flexible and uniform logging mechanism is available for consistent command-line and log-file output from within the whole analysis.

Finally, a set of general-purpose algorithms is included in PXL:
a node-based automatic layout, which is, e.g.,\ used in VISPA to display particle decay trees;
an automated reconstruction of all possible permutations of decay trees~\cite{autoprocess};
and the sorting as well as filtering of object collections.

For the longterm maintenance of the PXL and VISPA code, a server based on the \verb+redmine+~\cite{redmine} management web application is set up,
providing tools for bug tracking, project planning and code versioning using \verb+mercurial+~\cite{mercurial}.
Code development for multiple platforms is supported; the correct functionality is ensured by the usage of continuous integration with automatic builds and the execution of unit tests to validate the code on all target platforms. 

The components of PXL provide the infrastructure to build complex analyses in an object-oriented design.
The objects for HEP and astroparticle physics in C++ and Python, including the concepts of user data, relations, containers and I/O scheme, are designed to fulfill the requirements of good performance and easy handling.

\section{Modular Analysis System}

The PXL module system provides the interfaces and runtime environment to create and run physics analyses with individual modules.
It is the underlying analysis system of the \emph{Analysis Designer} in VISPA, but it can also be used independently of the VISPA development environment.
In the analysis, data are handed through a chain of modules. Each module can have multiple sinks for data input and sources for data output.
The modules contain the algorithms to process the data and steer it to different sources. This implies that the execution of the modules is based on the flow of the data through the analysis chain.

\begin{figure}[h!]
\begin{center}
\includegraphics[width=0.8\textwidth]{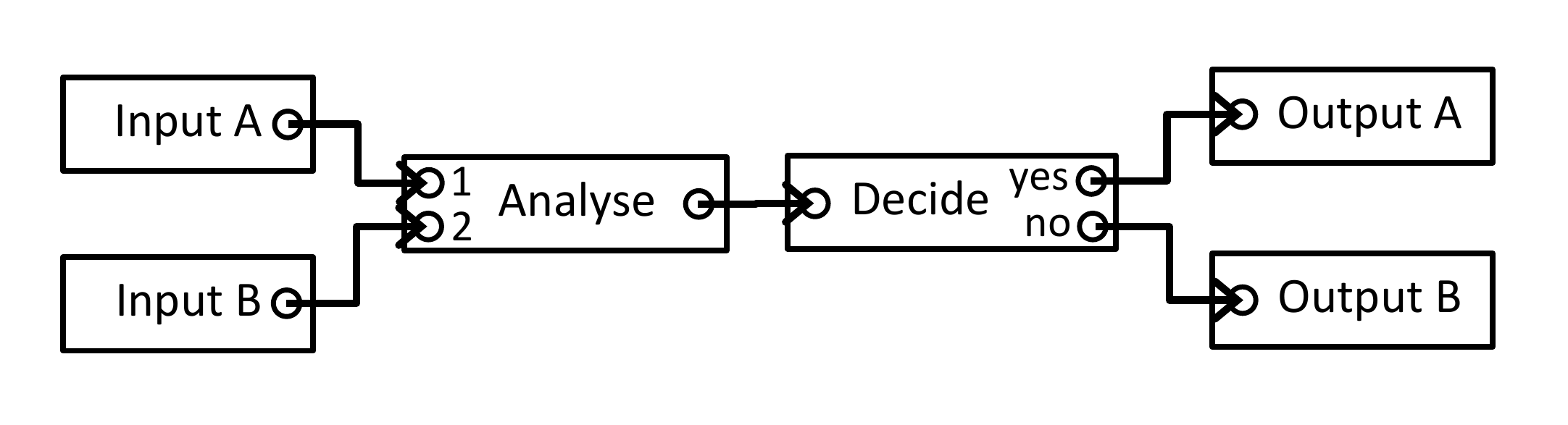}
\end{center}
\caption{Exemplary module chain.
Data are read in by the `Input A'  and `Input B' modules and are sent to the next module `Analyse'.
A `Decide' module splits the data stream, which is finally written into two different output files.}
\label{fig:modulechain}
\end{figure}

An example of a module chain is depicted in Figure~\ref{fig:modulechain}. The `Input A'  and `Input B' modules trigger the execution by reading a data chunk and passing it to the next module, an `Analyse' module that may, e.g.,\ run an algorithm on the data and add information to it. In the next module, labeled `Decide', analysis logic can be implemented by either sending the data to the `yes' or the `no' source and thereby deciding to which output module the data go.
In addition to the possibilities demonstrated in the figure, each sink can receive input from multiple sources.
The processing of the current data object stops if a module does not pass the object to any of its sources or if the source is not connected to another module.
By default, all modules able to start an analysis chain, e.g.,\ input or generator modules, are run in a single loop over the data.
Alternatively, by assigning a run index, they can be grouped into several loops over the data executed sequentially.
The input modules are able to read a list of files.
The use of more than one input module is in particular useful if categories of data are to be treated differently in the module chain.

For an appropriate overall turn-around time of the analysis development, it is important to be able to optimize computing-intensive tasks for high code performance and less intensive tasks for the minimum time spent by the user for code development.
One crucial design choice is therefore the use of two programming languages in the module system.
Users can freely choose to program their modules in C++ or Python and combine them in the analysis chain.
C++ is the de facto standard in HEP and astroparticle physics~\cite{cmsframework, athena, augerframework, root}.
Python is a widely used scripting language in science~\cite{root, scipy}, which
is often used to glue together code from different programming languages like C and Fortran~\cite{swig}.
While C++ code can be highly optimized for performance, Python as a scripting language allows fast and simple implementation of analysis logic.
A fast start into analysis using Python is well suited for beginners, while fast code prototyping in Python as well as high performance C++ coding can be achieved by experienced users.

By the use of C++ and Python, the module system is designed to be interfaced to a large variety of analysis- and experiment-related software.
External software is accessible through their C++ or Python interfaces from any analysis module.
This includes plotting packages (e.g.\ \verb+ROOT+~\cite{root}, \verb+matplotlib+~\cite{matplotlib}),
statistical analysis tools (e.g.\ \verb+RooStats+~\cite{roostats}, \verb+RooFit+~\cite{roofit}) and math/algebra packages (e.g.\ \verb+SciPy+~\cite{scipy}, \verb+NumPy+~\cite{numpy}).
Furthermore, interfaces or converters for data input from various experiments (CMS, ATLAS, Pierre Auger, ILC, ...) and file formats (LHE\cite{lhe}, CSV~\cite{CSV}) have been implemented.

The data passed through modules along the chain can be any derivative of a PXL class.
For physics analyses in HEP, typically, a container object representing a particle collision event holding a set of observed particles is used.
For astroparticle physics analyses a container object carrying a dataset of ultra-high energy particles is used.

All analysis modules implement an abstract interface provided in the module system.
It defines the access methods to data sinks, data sources, and module options; it also defines the methods executed on each data package before, during and after the execution of the analysis.
The module interface is identical in C++ and Python, thus simplifying the conversion of fast-prototyped Python modules into high-performance C++ modules.
VISPA is delivered with a set of predefined modules for file input/output and module skeletons for analysis logic (generate, decide, switch and analyse),
which are typically combined to an analysis with user-defined modules.

The analysis structure, including the modules as well as their connections, is stored in an object of the class \emph{Analysis}.
This class provides the functionality to store/retrieve an analysis setup to/from an XML file.
The class \emph{Analysis} can be used by two different executables: \emph{pxlrun}, which performs command line execution of analyses, and the graphical environment of VISPA, which allows the visual design and execution of analyses.
The full analyses including modules and data can be exported into an archive file from the VISPA environment, which can be used to exchange analyses with others and for batch processing.

The definition of a common interface for the exchanged data (e.g.\ Event for HEP) ensures the exchangeability of data between users.
The common interfaces of the modules and the \emph{Analysis} ensure the exchangeability and reusability of modules and analyses.
In particular, common modules can be shared within analysis groups, thereby facilitating teamwork.

The flow-based design is particularly well-suited for HEP analyses where particle collision event data are processed.
By analyzing the particles and other information in an event, decisions are typically taken whether to select certain events as depicted in Figure~\ref{fig:modulechain}.

A specific example for astroparticle physics analyses in VISPA are studies of extragalactic magnetic fields through ultra-high energy cosmic rays with the Pierre Auger Observatory~\cite{uhecr}.
Due to the experiment-independent format, VISPA can be used to transfer analyses from experiment to experiment. The analyses of Pierre Auger data constraining extragalatic magnetic fields can in principle be directly applied to a dataset from other cosmic ray experiments.

\section{Using the Components of the Development Cycle}

The process of analysis development and application in VISPA is centered around a graphical representation of the analysis in the \emph{Analysis Designer}.
This supports the analyzer in designing a well-structured and modular analysis and at the same time helps to structure his analysis work.
The main analysis tasks, like browsing input or output data in the \emph{Data Browser} or editing module code, are accessible via double click on the corresponding module in the graphical representation of the analysis.
The analysis execution can also be launched from the \emph{Analysis Designer}.

The \emph{Analysis Designer} is based on the implementation of modular physics analysis described in the previous section.
It visualizes the data flow between the modules of an analysis and provides access to any given module or parameter as sketched in Figure~\ref{fig:vispawindow}.
Modules can be added to the analysis from a list of predefined or custom C++ and Python modules.
The data flow is controlled by connecting sink and source ports of modules via drag and drop.
Module parameters can be modified in a \emph{Property View} that displays a table of parameter names and associated values.
While the complete analysis handling and the design of the analysis data flow can be performed visually, the textual programming of the actual modules is performed in the user's favorite editor.
For Python modules, the editor opens up with double clicking on a module.
Analyses can be executed from the \emph{Analysis Designer} via a button.
The command line output of the analysis execution can then be monitored in a separate section of the window.
The analysis execution happens in a separate process to keep the VISPA environment fully functional during execution.

The input and output data of an analysis can be explored within the same environment in the \emph{Data Browser}.
It visualizes all data in a PXL file, in particular decay trees of objects with relations, e.g.,\ particles, and gives access to all contained information in a \emph{Property View} as sketched in Figure~\ref{fig:vispawindow}.
The views of the \emph{Data Browser} represent data focused on its structure and full information coverage, as opposed to event displays which focus rather on the visualization in the context of detector geometries.
The two main purposes of the \emph{Data Browser} are the understanding of the input data and analysis debugging.
By browsing the input data, all the available information can be identified.
By browsing the output data of intermediate or final analysis steps, the functionality of each module, which may filter the data or add information to it, can be checked.

Finally, dialogs to execute the analysis or to send it to a batch system are provided.
For the submission of batch jobs, a dedicated plug-in is provided by VISPA, the \emph{Batch Manager}.
The \emph{Batch Manager} plug-in allows to configure batch jobs where input parameters of analyses may be iterated.
Hereby, one can execute slightly modified versions of a single analysis, a common use case in physics analysis.
The jobs may be sent to either a plug-in for multicore application on a single machine (\emph{Local Batch Manager})
or to a plug-in for submission to a Condor or Grid computing cluster (\emph{Condor/Grid Batch Manager}),
providing full flexibility for the execution of analyses from a single computer to large scale computing.

\section{Example Applications}

The applications of VISPA range from rather simple use cases to complex analyses.
In this section we provide a few examples for such cases with different complexity which are performed with VISPA.

The \verb+EasyROOTplot+~\cite{EasyRootPlot} modules are a collection of four plotting modules and a text file parser which takes text files in Comma Separated Value (CSV)~\cite{CSV} format as input.
They represent a plotting tool based on \verb+ROOT+~\cite{root}, which can be steered graphically in the \emph{Analysis Designer}.
An input module and plotting modules are plugged together, and parameters can be configured.
The text file parser converts the values found in the CSV file into an N-dimensional vector that is used as input for the plotting modules.
It allows the configuration via the \emph{Property View} (Fig.~\ref{fig:vispawindow}) to adapt to differently formatted files such as different separators, or data values stored in rows instead of columns. The plotting modules then take these N-vectors to produce plots, such as graphs or histograms. The variables to be plotted can be chosen via the \emph{Property View} and are identified by line headings in the CSV file or by identifier numbers generated at parsing time. The output can be chosen via a file save dialog and can be any picture format supported by \verb+ROOT+. For further customization such as adapting the plotting style, the user can modify the corresponding Python code lines marked within the \verb+EasyROOTplot+ modules. 

Another application of VISPA is browsing of Monte Carlo (MC) generator output in the form of the Les Houches Event (LHE)~\cite{lhe} data format.
The LHE format contains a list of particles with mother-daughter relations representing a particle production/decay chain.
In VISPA one can combine an LHE input module~\cite{lheinput} and a PXL output module to convert the LHE objects to PXL objects.
The resulting PXL file can be inspected with the \emph{Data Browser}.
In this way one can validate MC generator output which, due to its complexity with many related particles, can be best understood graphically and in an interactive manner.
The zooming functionality of the graphical views in VISPA helps to understand large particle decay chains at both detailed and large-scale level.

A plug-in related to the \emph{Data Browser} is the \emph{Event Editor}. This plug-in can be used, e.g.,\ to graphically
create a parton scattering process similar to a Feynman diagram. Particles can be selected from a list by
drag and drop and can be connected to build mother-daughter relations between them. The diagram
can be stored as an \emph{Event} of the PXL physics library or exported to an image file. Such diagrams are used as templates
,e.g.,\ for automated reconstruction of decay trees from final state particles as implemented in a
specialized module~\cite{autoprocess}. Typical applications at the LHC are the reconstruction of $Z$ or $W$ bosons, top quarks or hypothetical particles from extensions of the Standard Model of particle physics.

\begin{figure}[h!]
\begin{center}
\subfigure[]{\includegraphics[width=0.57\textwidth]{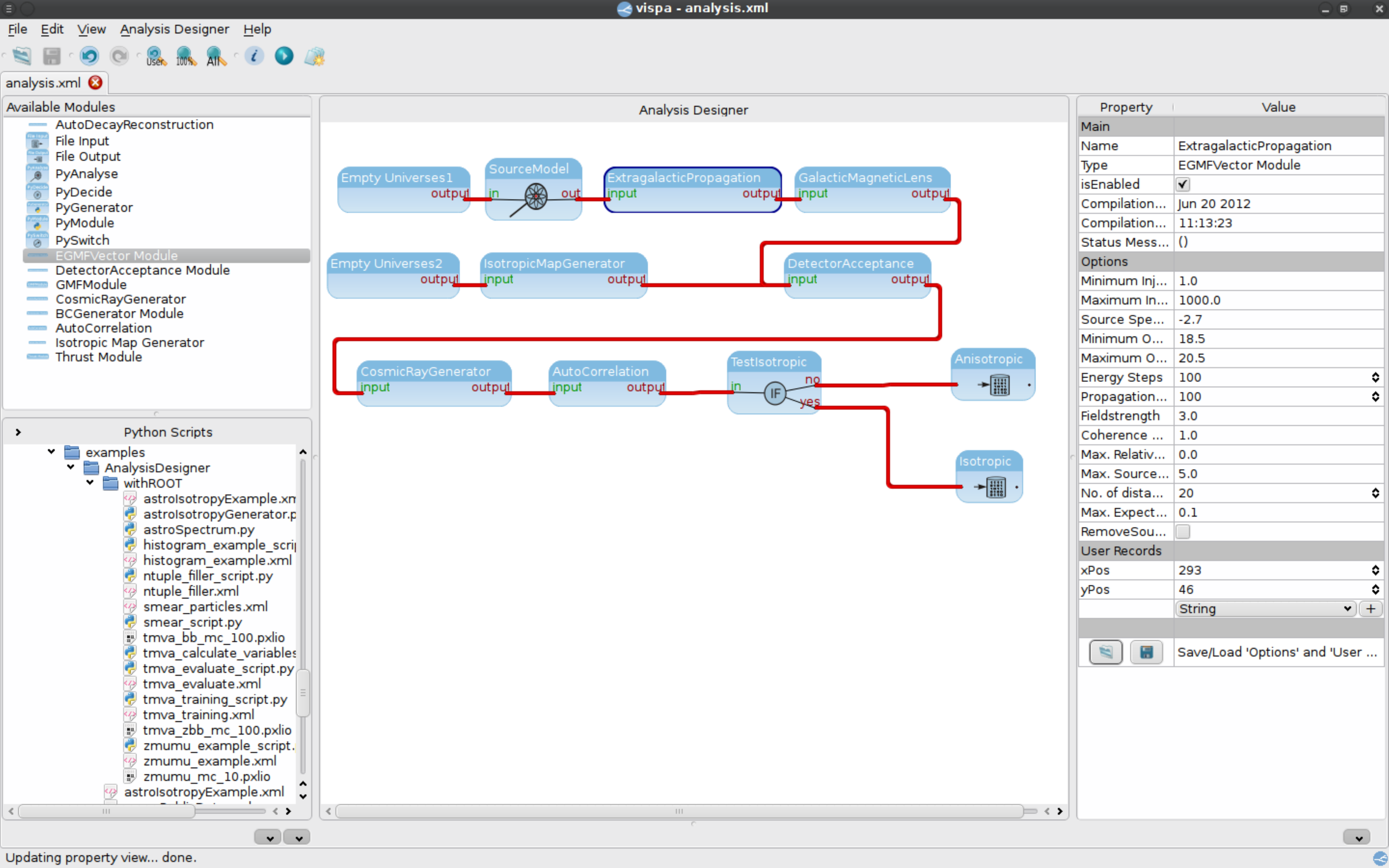}}
\subfigure[]{\includegraphics[width=0.57\textwidth]{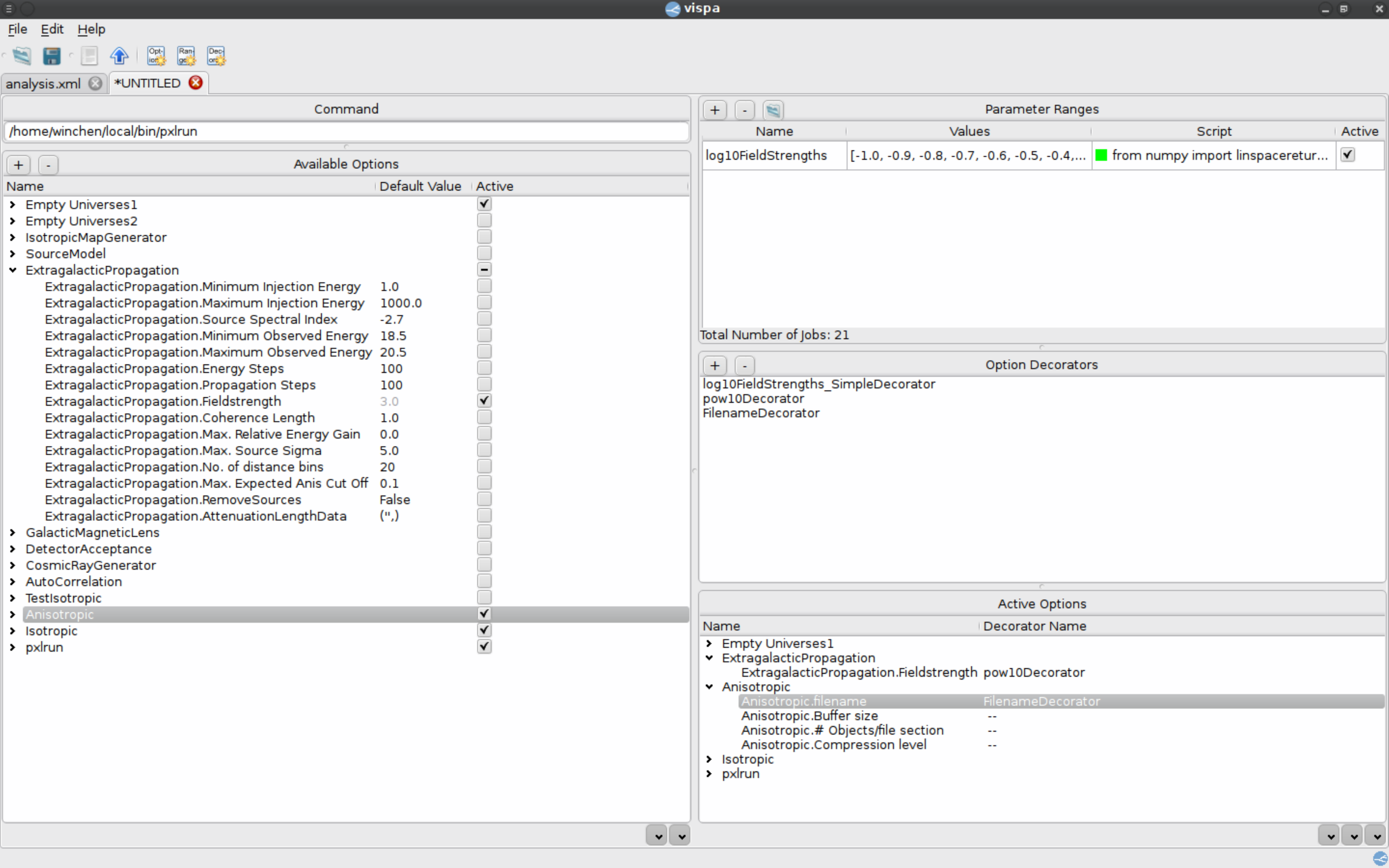}}
\subfigure[]{\includegraphics[width=0.57\textwidth]{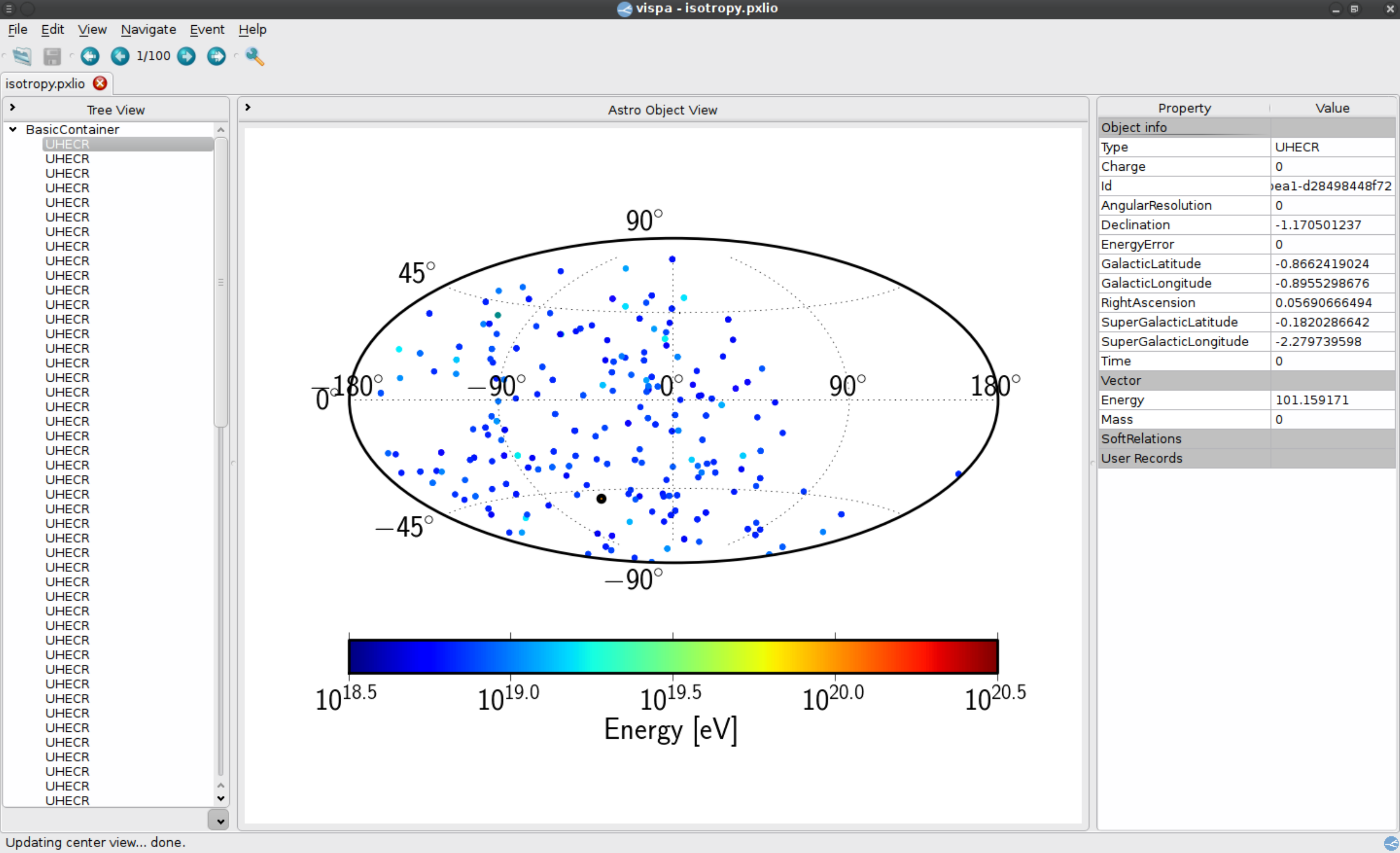}}
\end{center}
\caption{
(a) Exemplary simulation chain of the cosmic ray generator \texttt{PARSEC} in the \emph{Analysis Designer}. Isotropically distributed cosmic rays and cosmic rays from point sources are generated and analyzed. The parameters of the extragalactic magnetic field model can be modified in the \emph{Property View} on the right-hand side.
(b) \emph{Batch Job Designer} (part of the \emph{Batch Manager} plug-in) to create multiple simulation jobs with varying strength of the extragalactic magnetic field.
(c) Resulting arrival directions of isotropically distributed cosmic rays in galactic coordinates within a detector field of view, opened in the \emph{Data Browser}.
}
\label{fig:screenshots}
\end{figure}

A more complex application of VISPA is the Parametrized Simulation Engine for Cosmic Rays (\verb+PARSEC+)~\cite{parsec}, a MC generator to model energy-dependent anisotropies in the UHECR arrival distribution.
Simulated datasets are generated based on
models of the source distribution and the propagation in extragalactic
space as well as in our galaxy. The
simulation code is separated into individual VISPA modules whose
parameters can be accessed in the GUI. In a first step empty \emph{BasicContainers}
are generated and filled with \emph{UHECRSources} according to a user
defined model of the source distribution.
Based on the source distribution, maps of the
probability to observe a UHECR with an energy E from a direction $(\phi,
\theta)$ including deflection in extragalactic magnetic fields are calculated.
The maps are stored as \emph{BasicNVectors}. In the next step,
these probability maps are transformed by multiplying the vectors with
precalculated matrices to account for deflections in the coherent galactic
magnetic field of the milky way. In the final step of the MC
generator, individual cosmic rays are generated from these probability maps.
The underlying models of the simulation can be modified by either
changing the parameters accessible through the \emph{Property View}
or by changing the complete module.
The used modules and parameters are stored in the containers as
\emph{UserRecords} to keep track of the settings.
Besides these advantages of the modular design and GUI access,
the \emph{Data Browser} is frequently used to quickly check the simulation
output in a typical run. Furthermore, the \emph{Batch Manager} allows for the
convenient mass production of UHECRs for large parameter scans.
Figure~\ref{fig:screenshots} shows screenshots of a typical \verb+PARSEC+ analysis cycle with
the \emph{Analysis Designer}, the \emph{Batch Manager} and the \emph{Data Browser}.

Another rather complex use case in high energy physics is the application of multivariate data analysis techniques. Here, VISPA enables the dynamic integration of external packages designed for multivariate data analysis into one common analysis workflow.
The PXL module system is used to interface the configuration of a corresponding package such as TMVA~\cite{tmva}. Sets of parameters used to configure a multivariate analysis package can be visualized and steered via the \emph{Property View}, which allows to modify the most important and most used options.
VISPA supports transparent and user-defined splitting of data streams into categories as used for training, testing and evaluation purposes. The data stream can be further divided into separate sub-samples of the available phase space, depending on the kinematics specific to particular physics processes. Categorizing data is particularly useful when several classifiers are needed, e.g.,\ to build super-discriminants of diversified multivariate classifiers or to chain several classifiers.
Technically, the classifier output can be conveniently stored as a \emph{UserRecord} in each \emph{Event}. Classifier output values using various configurations can be stored consecutively in the same \emph{Event}. This feature allows to compare the performance of diversified classifiers and configurations. In this way, an optimization of the used classifiers with respect to consecutive analysis steps is feasible. An important example application is the selection of the classifier resulting in the best expected sensitivity of a particular analysis.

An example for an experiment-specific plug-in that integrates experiment-specific tasks of the analysis flow into the VISPA development environment is the \verb+ConfigEditor+~\cite{configbrowser}.
It is a widely used tool for inspecting the configuration of workflows in the CMS experiment and shares most of its GUI implementation with the \emph{Analysis Designer}.
For VISPA analyses the \verb+ConfigEditor+ is used for the analysis step in which objects relevant to the analysis are configured and selected from a large variety of reconstructed objects available from the experiment.
For CMS, it is generally useful for debugging of configurations of any step of the data processing chain of the experiment: trigger, reconstruction, event generation and simulation.

\section{Conclusions}

The VISPA graphical development environment provides the tools to iterate through a full analysis cycle.
For designing and programming of physics analyses, three different software paradigms are combined.
First, physics analyses are developed using visual programming.
The development process combines visual design of the analysis flow in the graphical user interface and textual programming of individual modules.
Second, in the context of flow-based programming, the data are handed through chains of modules that apply different algorithms implementing the analysis logic.
Third, algorithms are designed with object-oriented textual programming based on the objects provided by PXL.
VISPA also provides the tools for local or distributed execution, and for validation of the results.

To ensure a flexible analysis design and to cover all steps of physics analyses in different fields of physics, the VISPA user interface has been designed as a plug-in-based framework.
This enables the extension of VISPA with more analysis-related tools and the application of VISPA in other fields of research besides HEP and astroparticle physics.
Because of its platform-independent implementation and its well-defined analysis and data interfaces, VISPA particularly facilitates teamwork in analysis groups. 

\acknowledgments

We wish to thank Benedikt Hegner for fruitful discussions and valuable comments on the manuscript.
For important contributions to the first version of PXL, we thank Steffen Kappler.
For fruitful discussions during the development phase of PXL, we thank Matthias Kirsch.
We thank Carsten Hof, Philipp Biallass and Holger Pieta for careful evaluation of the PXL I/O system.
We also thank Anna Henrichs for providing a PXL interface to the data format of the ATLAS experiment.
For the creation of the user manual for PXL, we thank Oxana Actis.
This work is supported by the Ministerium f\"ur Wissenschaft und Forschung, Nordrhein-Westfalen, the Bundesministerium f\"ur Bildung und Forschung (BMBF), and the Helmholz Alliance "Physics at the Terascale".
T. Winchen gratefully acknowledges funding by the Friedrich-Ebert-Stiftung. 


\bibliography{vispapaper}{}
\bibliographystyle{JHEP}

\end{document}